\documentclass[referee]{aa}
\usepackage{graphicx}
\usepackage{natbib}
\usepackage{url}
\usepackage{amsmath}
\bibpunct{(}{)}{;}{a}{}{,}

\begin{document}

\title{Expectation Maximization for Hard X-ray Count Modulation Profiles}

   \author{Federico Benvenuto\inst{1}
          \and
          Richard Schwartz\inst{2}
          \and
          Michele Piana\inst{1,3}
          \and
          Anna Maria Massone\inst{3}}

   \institute{	Dipartimento di Matematica, Universit\`a di Genova, via Dodecaneso 35, 16146 Genova, Italy \and
			  	Catholic University and Solar Physics Laboratory, Goddard Space Flight Center, code 671, Greenbelt, MD 20771 USA \and
 				 CNR - SPIN, via Dodecaneso 33, I-16146 Genova, Italy
			}
   \date{}

\abstract{This paper is concerned with the image reconstruction problem when the measured data are solar hard X-ray modulation profiles obtained from 
the \textit{Reuven Ramaty High Energy Solar Spectroscopic Imager (RHESSI)} instrument.}
{Our goal is to demonstrate that a statistical iterative method classically applied to the image deconvolution problem is very effective when utilized for the analysis of count modulation profiles in solar hard X-ray imaging based on Rotating Modulation Collimators.}
{The algorithm described in this paper solves the maximum likelihood  problem iteratively and encoding a positivity constraint into the iterative optimization scheme. The result is therefore a classical Expectation Maximization method this time applied not to an image deconvolution problem but to image reconstruction from count modulation profiles. The technical reason that makes our implementation particularly effective in this application is the use of a very reliable stopping rule which is able to regularize the solution providing, at the same time, a very satisfactory Cash-statistic (C-statistic).}
{The method is applied to both reproduce synthetic flaring configurations and reconstruct images from experimental data corresponding to three real events. In this second case, the performance of Expectation Maximization, when compared to Pixon image reconstruction, shows a comparable accuracy and a notably reduced computational burden; when compared to CLEAN, shows a better fidelity with respect to the measurements with a comparable computational effectiveness.}
{If optimally stopped, Expectation Maximization represents a very reliable method for image reconstruction in the \textit{RHESSI} context when count modulation profiles are used as input data.}


\authorrunning{Benvenuto et al.}
\titlerunning{Expectation Maximization for hard X-ray modulation profiles}

\maketitle

\section{Introduction}\label{intro}

Expectation Maximization (EM) \cite{delaru77} is an iterative algorithm that addresses the maximum likelihood problem when 1) the relation between the unknown parameter (or set of parameters) and the measured data is linear; 2) the data are drawn from a Poisson distribution; 3) the unknown parameter satisfies the positivity constraint. EM represents a generalization of the image restoration method introduced by Lucy and Richardson \cite{lu74} in the case of deconvolution problems typical of focused astronomy and is successfully applied to several reconstruction problems in optics, microscopy and medical imaging. The present paper applies EM for the first time to the hard X-ray count modulation profiles measured by the nine Rotating Modulation Collimators (RMCs) mounted on the Reuven Ramaty High Energy Solar Spectroscopic Imager (\textit{RHESSI}) \cite{lietal02}. More specifically, we show here that, when combined with an optimal stopping rule for the iterative process, EM provides reliable reconstructions with notable computational effectiveness.

The \textit{RHESSI} imaging concept translates to a solar context the rotational modulation synthesis first introduced for non-solar observations \cite{huetal02}. In \textit{RHESSI}, a set of nine rotating collimators, characterized by nine pairs of grids with nine different pitches, time-modulates the incoming photon flux before it is detected by the nine corresponding Ge crystals. The resulting signal is a set of nine time series representing the count evolution provided by each collimator-detector system at different time bins. Therefore, these count modulation profiles represent the temporal or rotation angle/phase variation of the count rates for each grid.

Since the transformation from the flux distribution on the image plane to the set of count modulation profiles is linear, the \textit{RHESSI} image reconstruction problem is the linear inverse problem of describing the flux distribution from the count modulation profiles. EM describes the observed data and the unknown as realizations of stochastic quantities and searches for the  flux distribution that maximizes the probability of the observation under the constraint that the pixel content must be positive. In fact, when the noise distribution on the data is Poisson this constrained maximum likelihood problem can be transformed into a fixed point problem whose solution is obtained iteratively, by means of a successive approximation scheme. From a theoretical viewpoint, the convergence properties of this algorithm when the number of iterations grows are not completely known. However, it is always observed in applications that stopping the procedure at some optimal iteration regularizes the reconstruction, thus preventing the occurrence of over-resolving effects or small-wavelength artifacts. In this paper, this optimized stopping rule is determined by utilizing the concept of constrained residual and by imposing that the empirical expectation value of this stochastic variable coincides with its theoretical expectation value. 

The plan of the paper is as follows. In Section 2 we describe the \textit{RHESSI} imaging concept in more detail and introduce the linear transformation modeling the data formation process. Section 3 contains the formulation of the iterative algorithm together with its stopping rule. Section 4 validates EM in the case of several synthetic modulation profiles simulated from plausible configurations of the flux distribution. Finally, in Section 5 we consider the application of EM to three sets of observed \textit{RHESSI} data. Our conclusions are in Section 6.

\section{\textit{RHESSI} count modulation profiles}
The \textit{RHESSI} imaging hardware is made of nine sub-collimators, each one consisting of a pair of separated grids in front of a hard X-ray / gamma-ray detector. In each sub-collimator the two grids are identical, parallel and characterized by a planar array of equally-spaced, X-ray-opaque slats separated by transparent slits. The grid pitches of the different collimators are arranged according to a geometric progression with factor $\sqrt{3}$ with detector 1 providing the maximum resolution power and minimum signal-to-noise ratio and detector 9 providing the minimum resolution power and maximum signal-to-noise ratio. \textit{RHESSI} rotates around its own axis with a period of around $4$ s and this rotation, combined with the presence of the grids, induces the modulation of the count rates. Therefore in this framework, there is no detector plane containing physical pixels (like in focused imaging) and here pixels are just a mathematical idealization in the image reconstruction process. If we describe the brightness distribution by means of the vector $f$ of dimension $N^2 \times 1$ (in this lexicographic ordering each vector index denotes one of the $N^2$ pixels), then the expected count modulation detected by sub-collimator $l$ is given by
\begin{equation}\label{forward-1}
g^{(l)}  = H^{(l)} f,
\end{equation}
where the vector $g^{(l)}$ has dimension $P^{(l)} \times 1$, $P^{(l)}$ is the number of time bins discretizing the time evolution of the modulation for detector $l$ and $H^{(l)}$, with dimension $P^{(l)} \times N^2$, is the matrix modeling the transformation from the image to the measurement space. The entries of $H^{(l)}$ can be interpreted in a probabilistic way as
\begin{equation}\label{forward-1}
H^{(l)}_{im} = A {\mathcal{P}}^{(l)}_{im} \triangle t_i
\end{equation}
where $A$ measures the detector area and ${\mathcal{P}}^{(l)}_{im}$ is the probability that a photon originating in pixel $m$ will be counted in the $i$-th time bin of detector $l$ during the time interval $\triangle t_i$. If the image analysis performed involves data collected by $M$ of the nine sub-collimators, then the overall signal formation process is described by
\begin{equation}\label{forward-3}
g = H f,
\end{equation}
where $g$ has dimension $L \times 1$, $L=\sum_{l=1}^{M} P^{(l)}$ and contains all the modulation profiles while $H$ has dimension $L \times N^2$ and represents the image operator mimicking the action of the telescope. The EM algorithm together with its optimal stopping rule provides an estimate of $f$ by means of an iterative regularized inversion of $H$.

\section{The EM algorithm}
Expectation Maximization is a statistical algorithm maximizing the probability that the data vector is a realization corresponding to a Poisson random vector $g$. In fact, in this case, the likelihood, i.e. the probability to observe $g$ from the model $Hf$, can be written as
\begin{equation}\label{likelihood-imaging}
P(g|f) = \prod_{i=1}^L e^{-(Hf)_i} \frac{(Hf)_{i}^{g_i}}{g_i !},
\end{equation}
where the ratio should be intended point-wise, element by element. We observe that maximizing this probability corresponds to minimizing the Cash statistic (C-statistic) \citep{ca79}
\begin{equation}\label{c-statistic-imaging}
C_{stat} (g,f) = \frac{2}{L} \sum_{i=1}^L g_i \log \frac{g_i}{(Hf)_i} + (Hf)_i - g_i.
\end{equation}
The likelihood maximizer is constrained to the set of positive solutions, i.e. the algorithm solves the constrained optimization problem
\begin{equation}
\arg\max_{f \geq 0} P(g|f) = \arg\min_{f \geq 0} C_{stat}(g,f),
\end{equation}
which can be transformed into a fixed-point problem solved by means of the successive approximation scheme
\begin{equation}\label{em-algorithm-imaging}
f_{k+1} = f_k \frac{H^{T} \left(\frac{g}{Hf_k}\right)}{H^T 1},
\end{equation}
with a positive (constant) initialization and where $1$ denotes the vector made of all unit entries. Since $H$ is ill conditioned, this iterative algorithm should be regularized by applying some stopping rule. To this aim we observe that the asymptotical behavior of equations (\ref{em-algorithm-imaging}) is such that either $f_k \rightarrow 0$ or
\begin{equation}\label{auxiliary}
\alpha_k = \frac{H^{T} \left(\frac{g}{Hf_k}\right)}{H^T 1}
\end{equation}
converges to $1$. This implies that, asymptotically, 
\begin{equation}\label{em-imaging-stopping-1}
z_k = \| f_k H^{T}\left(1 - \frac{g}{Hf_k}\right) \|^2
\end{equation}
tends to zero and therefore a reasonable stopping rule for EM in the \textit{RHESSI} case is
\begin{equation}\label{em-imaging-stopping}
z_k = E(z_k),
\end{equation}
where $E(z_k)$ denotes the expectation value of $z_k$.

\section{Numerical validation}
\begin{figure}[pht]
\begin{center}
\includegraphics[width=1.\textwidth]{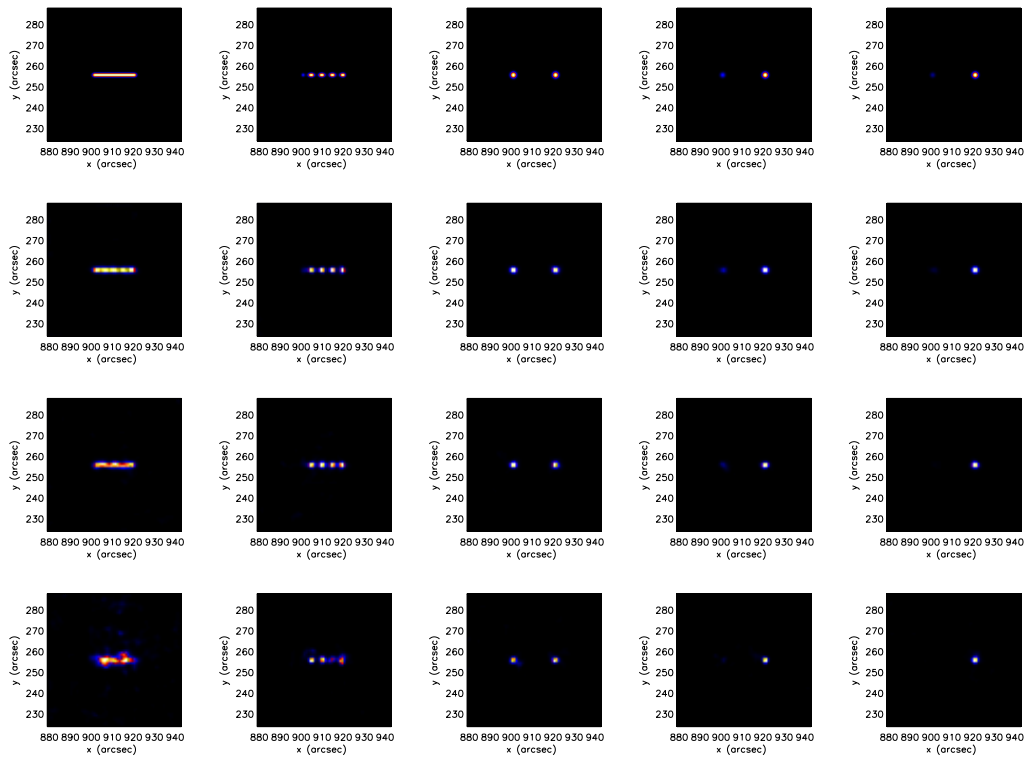}
\caption{Validation of EM with synthetic data. First row: the simulated configurations. Rows 2 through 4: reconstructions provided by EM corresponding to count modulation profiles characterized by three different levels of statistics (second row: average of 100000 counts per detector; third row: average of 10000 counts per detector; fourth row: average of 1000 counts per detector. }
\label{fig:simulated-reconstructions}
\end{center}
\end{figure}

In order to assess the reliability of EM we setup a validation test based on the following process:
\begin{enumerate}
\item Five different configurations of the flaring region were invented (see Figure \ref{fig:simulated-reconstructions}), first row, characterized by very different topographical and physical properties (e.g., size, position, number and distance of disconnected components, relative intensity of the components). Specifically, the original configurations are: a line source with constant density along the line (case A); a line source with intensity varying along the line, i.e. four compact sources and a weak one, all sources being aligned (case B); two Gaussian sources with flux ratio equal to 1 (case C); two Gaussian sources with flux ratio equal to 5 (case D); two Gaussian sources with flux ratio equal to 10 (case E);
\item For each flaring configuration, three different synthetic sets of count modulation profiles were realized, characterized by three different levels
of statistics (low, medium, high). Operationally, matrix $H$ was applied to the simulated map, the resulting count expected values at each time bin were scaled with three different values in order to simulate three different levels of statistics (an average of 1000 counts per detector for the low level, 10000 for the medium level and 100000 for the high one);
\item EM was applied to each one of the resulting 15 data sets in order to reconstruct the images;
\item A set of routines was applied, for the quantitative assessment of the algorithm performance. These routines compute specific physical and geometrical parameters in the images, that are particularly significant for the different configurations, and compare the values with the corresponding ground truth values in the simulation maps of Figure \ref{fig:simulated-reconstructions}, first row. 
\end{enumerate}
Figure \ref{fig:simulated-reconstructions}, rows 2 through 4, contains the reconstructions provided by EM for the three different levels of statistics and using the count modulation profiles provided by all nine \textit{RHESSI} detectors. The assessment routines are applied to these maps and compute the following parameters:

Case A (line source with constant intensity):
\begin{itemize}
\item A1: orientation (ground truth: $0$ deg).
\item A2: number of reconstructed sources (ground truth: 1). While reconstructing a line source, most (if not all) imaging methods tend to break it up into a set of compact sources (this, particularly, occurs at low levels of statistics). The routine computes the number $K$ of intersection knots between the reconstructed line profile and the straight line with the same orientation passing through the Full Width at Half Maximum (FWHM). The number of reconstructed sources is counted as $K/2$.
\item A3: length (ground truth: $20.2$ arcsec). The routine computes the FWHM in the direction of the orientation line.
\item A4: width (ground truth: $1.35$ arcsec). The routine computes the FWHM in the direction orthogonal to the orientation line.
\end{itemize}

Case B (line source with intensity varying along the line)
\begin{itemize}
\item B1: orientation (ground truth: $0$ deg).
\item B2: number of reconstructed sources at FWHM (ground truth: 4). Computed as in case A2.
\item B3: length (ground truth: $16.9$ arcsec). Computed as in case A3.
\item B4: width (ground truth: $1.35$ arcsec). Computed as in case A4.
\end{itemize}

Case C (two sources with flux ratio 1)
\begin{itemize}
\item C1: position of the first reconstructed source (ground truth: $0$ arcsec). The routine computes the distance between the peak of the first reconstructed source and the corresponding simulated one.
\item C2: position of the second reconstructed source (ground truth: $0$ arcsec). The routine computes the distance between the peak of the second reconstructed source and the corresponding simulated one.
\item C3: separation of the reconstructed sources (ground truth: $20$ arcsec). The routine computes the distance between the two peaks.
\item C4: orientation of the separation line (ground truth: $0$ deg). The routine computes the orientation of the line passing through the two reconstructed peaks.
\item C5: flux ratio (ground truth: 1). For each simulated source, the routine computes the disk centered in correspondence with the peak and with radius such that 99\% flux is within the disk. Then, in the reconstructed image, the routine computes the fluxes contained in the two disks and makes the ratio.
\end{itemize}

For Case D (two sources with flux ratio 5) and Case E (two sources with flux ratio 10) the routines compute the same parameters as in Case C. 

We performed this analysis by comparing the parameters obtained by EM with the original simulation parameters and with the ones obtained by different imaging algorithms, namely: Pixon \cite{pu95,meetal96}, CLEAN \cite{ho74}, Maximum Entropy \cite{boetal06}, a forward-fit algorithm for visibilities and uv\_smooth \cite{maetal09} (for all algorithms we used all nine \textit{RHESSI} detectors).
In Table \ref{pixon-em-simulated} we reported just the results provided by Pixon since, among the methods using modulation profiles as input, provided among the best results. The Pixon algorithm models the source as a superposition of circular sources (or {\textit{pixons}}) of different sizes and parabolic profiles and looks for the one that best reproduces the measured modulations from the different detectors. This technique is generally considered as the most reliable one in providing the most accurate image photometry \cite{depe09}, but at the price of a very notable computational burden. In this experiment we configured the Pixon algorithm in Solar SoftWare (SSW) according to an optimized procedure based on heuristic arguments.

\begin{table}
\begin{center}
\begin{tabular}{c|c|c||c|c||c|c||c}
\hline
 & EM & Pixon  & EM & Pixon & EM & Pixon & g.t. \\
 & $10^5$ & $10^5$  & $10^4$ & $10^4$ & $10^3$ & $10^3$ &  \\
 \hline
 A1 & -0.06 & -0.1 & -0.41 & -0.61 & 3.33 & 3.84 & 0.0 \\
 A2 & 1 & 1 & 3 & 2 & 2 & 2 & 1 \\
 A3 & 18.86 & 19.81 & 17.84 & 18.31 & 13.51 & 14.47 & 20.2 \\
 A4 & 2.08 & 3.04 & 1.94 & 2.17 & 3.45 & 4.27 & 1.35 \\
 \hline \hline
 B1  & -0.04 & -0.10 & -0.06 & 0.01 & -0.96 & 0.11 & 0.0 \\
 B2 & 4 & 4 & 4 & 4 & 3 & 3 & 4 \\
 B3  & 16.41 & 17.82 & 16.35 & 16.58 & 15.92 & 17.13 & 16.90 \\
 B4  & 2.05 & 3.79 & 2.17 & 2.81 & 1.86 & 3.10 & 1.35 \\
 \hline \hline
 C1  & 0.71 & 0.71 & 0.71 & 0.71 & 0.71 & 0.71 & 0.0 \\
 C2  & 0.71 & 0.71 & 0.71 & 0.71 & 0.71 & 0.71 & 0.0 \\
 C3  & 19.03 & 22.02 & 19.00 & 19.00 & 20.00 & 19.00 & 20.0 \\
 C4  & 3.01 & -2.6 & 0 & 0 & 0 & 0 & 0.0\\
 C5 & 1.01 & 1.01 & 1.01 & 1.02 & 0.99 & 1.00 & 1.0 \\
 \hline \hline
 D1  & 0.71 & 1.58 & 0.71 & 1.58 & 0.71 & 3.54 & 0.0 \\
 D2  & 0.71 & 1.58 & 0.71 & 1.58 & 0.71 & 3.54 & 0.0 \\
 D3  & 20.00 & 18.00 & 19.03 & 19.00 & 20.03 & 24.03 & 20.0 \\
 D4  & 0 & 0 & -3.01 & 0 & -2.86 & -2.39 & 0.0 \\
 D5 & 4.73 & 5.14 & 4.99 & 5.57 & 7.56 & 11.53 & 5.0 \\
 \hline \hline
E1 & 0.71 & 1.58 & 0.71 & 7.52 & 3.54 & 9.93 & 0.0 \\
E2  & 0.71 & 1.58 & 0.71 & 7.52 & 3.54 & 9.93 & 0.0 \\
E3 & 19.00 & 19.00 & 19.00 & 28.02 & 23.00 & 27.66 & 20.0 \\
E4 & 0 & 0 & 0 & -2.05 & 0 & -12.53 & 0.0 \\
E5 & 10.02 & 11.65 & 16.15 & 16.96 & 29.68 & 42.69 & 10.0 \\
 \hline \hline
\end{tabular}
\caption{Assessment of EM performances in the case of synthetic data; $10^5$, $10^4$ and $10^3$ (counts per detector) indicate the three different levels of statistics considered in the test. The comparison is made with the parameters characterizing the simulated configurations (ground truth, last column). Case A concerns the line source with constant intensity (Figure \ref{fig:simulated-reconstructions}, first row, first panel): orientation (A1, deg); number or reconstructed sources (A2); length (A3, arcsec); width (A4, arcsec). Case B concerns the line source with intensity varying along the line (Figure \ref{fig:simulated-reconstructions}, first row, second panel): orientation (B1, deg); number of reconstructed sources (B2); length (B3, arcsec); width (B4, arcsec). Case C concerns two sources with flux ratio 1 (Figure \ref{fig:simulated-reconstructions}, first row, third panel): position of the first reconstructed source (C1, arcsec); position of the second reconstructed source (C2, arcsec); separation (C3, arcsec); orientation of the separation line (C4, deg); flux ratio (C5). Case D (Figure \ref{fig:simulated-reconstructions}, first row, fourth panel) and Case E (Figure \ref{fig:simulated-reconstructions}, first row, fifth panel) as for Case C but the flux ratio 5 and 10, respectively.}
\label{pixon-em-simulated}
\end{center}
\end{table}

\section{Application to real observations}
\begin{figure}[pht]
\begin{center}
\includegraphics[width=1.\textwidth]{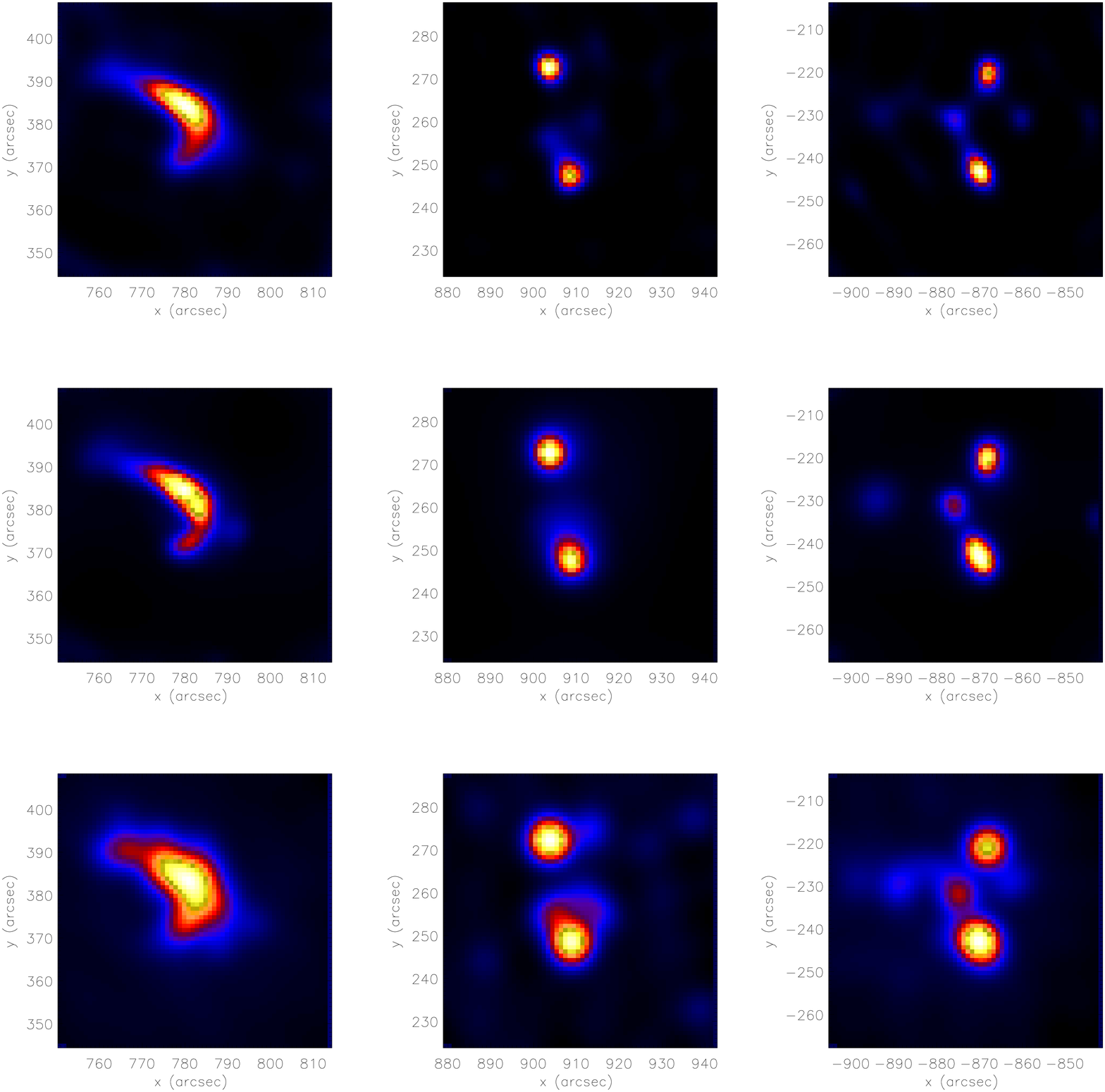}
\caption{Performance of EM in the case of real data observed by \textit{RHESSI}. First row: EM reconstructions; second row: Pixon reconstructions; third row: CLEAN reconstructions. First Column: 15 April 2002 event; second column: 20 February 2002 event; third column: July 23 2002 event.}
\label{fig:real-reconstructions}
\end{center}
\end{figure}

We applied EM to \textit{RHESSI} observations recorded in correspondence with three real flaring events. We considered the flare of April 15, 2002 in the time interval 00:06:00 -- 00:08:00 UT and in the energy interval 12 -- 14 keV; the flare of February 20, 2002 in the time interval 11:05:58 -- 11:06:41 UT and in the energy interval 25 -- 30 keV; the flare of July 23, 2002 in the time interval 00:30:00 -- 00:32:00 UT and in the energy interval 100 -- 300 keV. In all cases detectors 3 through 8 are used for the observations. The reasons for this choice are as follows: detector 2 has been characterized by malfunctions since the beginning of the \textit{RHESSI} mission; detector 1 is characterized by a very small signal-to-noise ratio while the coarse information carried by detector 9 is not crucial for the reconstruction of these events. Figure \ref{fig:real-reconstructions} compares EM reconstructions with the ones provided by Pixon and CLEAN, while Table \ref{em-pixon-real} contains the corresponding C-statistic for the six detectors employed in the analysis. In these experiments CLEAN parameters have been chosen according to optimized heuristic recipes \cite{depe09}. According to these results,
EM and Pixon are characterized by values of the C-statistic almost systematically close to 1 (and significantly smaller than the ones provided by CLEAN). This means the EM and Pixon can reproduce the data with comparable accuracy (significantly better than the one achieved by CLEAN), although EM reduces the computational time of up to a factor 4. Indeed, for April 15, 2002, the computational time is around 100 sec for EM and more than 400 sec for Pixon; for February 20, 2002, the computational time is around 50 sec for EM and almost 220 sec for Pixon; for July 23, 2002, the computational time is around 125 sec for EM and more than 460 sec for Pixon. This same 4 to 1 scaling holds true independently of the kind of hardware used for the tests. We also observe that in the Pixon implementation we used for these experiments the computations of the time profiles and back projections are done more efficiently by using optimized combinations of the spatial variation as spatial sine and cosine patterns (annsec, annular-sector, implementation). Implementing EM according to this same representation will improve the computational gain provided by this algorithm of another factor 5.

\begin{table}
\begin{center}
\begin{tabular}{c}
April 15 2002
\end{tabular}
\end{center}
\begin{center}
\begin{tabular}{c|c|c|c|c|c|c}
\hline
 & 3 & 4 & 5 & 6 & 7 & 8 \\
 \hline
EM & 1.360 & 1.384 & 1.207 & 2.207 & 2.312 & 5.302 \\
 \hline
Pixon & 1.459 & 1.507 & 1.486 & 2.302 & 2.464 & 5.111 \\
 \hline
 CLEAN & 14.50 & 13.98 & 13.86 & 27.65 & 41.07 & 77.79 \\
 \hline
 \end{tabular}
 \end{center}
 \begin{center}
 \begin{tabular}{c}
 February 20 2002
 \end{tabular}
 \end{center}
\begin{center}
\begin{tabular}{c|c|c|c|c|c|c}
\hline
 & 3 & 4 & 5 & 6 & 7 & 8 \\
 \hline
EM & 1.093 & 1.104 & 1.037  & 1.157 & 1.056 & 0.882 \\
 \hline
Pixon & 1.212 & 1.250 & 1.154 & 1.167 & 1.302 & 1.106 \\
 \hline
 CLEAN & 1.725 & 1.695 & 1.721 & 2.194 & 2.505 & 3.249 \\
 \hline
 \end{tabular}
 \end{center}
 \begin{center}
 \begin{tabular}{c}
July 23 2002
\end{tabular}
\end{center}
\begin{center}
\begin{tabular}{c|c|c|c|c|c|c}
\hline
 & 3 & 4 & 5 & 6 & 7 & 8 \\
 \hline
 EM & 1.080 & 0.973 & 1.220 & 1.341 & 1.690 & 1.837 \\
 \hline
Pixon & 1.176 & 1.007 & 1.224 & 1.340 & 1.702 & 1.797 \\
\hline
CLEAN & 5.302 & 5.326 & 4.711 & 5.756 & 3.908 & 9.878 \\
\hline
\end{tabular}
\end{center}
 \caption{Performance of EM in the case of real data observed by \textit{RHESSI}: comparisons of C-statistic provided by EM, Pixon and CLEAN.}
 \label{em-pixon-real}
 \end{table}

\section{Conclusions}
This papers shows that Expectation Maximization can be effectively applied to reconstruct hard X-ray images of solar flares from the count modulation profiles recorded by the \textit{RHESSI} mission. This method is an iterative likelihood maximizer with a positivity constraint, that explicitly exploits the fact that the noise affecting the measured data has a Poisson nature. We have utilized an optimal stopping rule that regularizes the algorithm, realizing an optimal trade-off between the C-statistic and the numerical stability of the reconstruction. We are aware that this test on C-statistic in not conclusive, since images affected by significant artifacts may reproduce the experimental data with great accuracy. However low C-statistic values coupled with the positivity constraint can be considered as a diagnostic of reliable reconstructions.

We have validated EM against synthetic count modulation profiles corresponding to challenging simulated configurations and characterized by three levels of statistics. Then we have applied the method against the \textit{RHESSI} observations of three flaring events and compared the reconstructions with the ones provided by Pixon and CLEAN. The results of these experiments show that EM combines a reconstruction fidelity (in terms of C-statistic) comparable with the one provided by Pixon (which, however, is much more demanding from a computational viewpoint) with a computational efficiency comparable with the one offered by CLEAN (which, however, predicts the count modulation profiles by means of a significantly worse C-statistic).
Our next step, which is currently under construction, will be to generalize this approach to the reconstruction of electron flux maps of the flaring region. Electron flux maps of solar flares can be already generated by hard X-ray count visibilities \cite{pietal07}. We are currently working at an EM-based approach to the reconstruction of electron images, where the input data are the count modulation profiles and the imaging matrix to invert accounts for both the effects of the bremsstrahlung cross-section and the Detector Response Matrix mimicking the projection from the photon to the count domain. The advantage of this approach with the respect to the visibility-based one should be that EM provides an analysis framework that is closer to the data, as we can model with a greater accuracy all of the detector effects.

\begin{acknowledgements}
The experiment with synthetic data in Section 4 has been conceived in collaboration with A. G. Emslie and G. H. Hurford, who are kindly acknowledged. This work was supported by the European Community Framework Programme~7, ``High Energy Solar Physics Data in Europe (HESPE)'', Grant Agreement No. 263086.
\end{acknowledgements}

\end{document}